\documentclass[conference]{IEEEtran}

\usepackage[utf8]{inputenc}
\usepackage[T1]{fontenc}

\usepackage{color}
\usepackage{amsmath}
\usepackage{amssymb}
\usepackage{booktabs}
\usepackage{dsfont}
\usepackage{listings}
\usepackage{manfnt}
\usepackage{subfig}
\usepackage{tabularx}
\usepackage{tikz}
\usepackage{graphics}
\usetikzlibrary{calc,decorations,decorations.pathreplacing,patterns,positioning}

\newtheorem{definition}{Definition}

\newcommand{\B}{\mathds{B}}
\newcommand{\bra}[1]{{\left\langle{#1}\right\vert}}
\newcommand{\ket}[1]{{\left\vert{#1}\right\rangle}}

\def\hang{\hangindent19pt}
\def\d@anger{\medbreak\begingroup\clubpenalty=10000
 \def\par{\endgraf\endgroup\medbreak} \noindent\hang\hangafter=-2
 \hbox to0pt{\hskip-\hangindent\dbend\hfill}\small}
\outer\def\danger{\d@anger}

\definecolor{dgreen}{rgb}{0,.4,0}
\definecolor{dblue}{rgb}{0,0,.7}
\definecolor{dkgreen}{rgb}{0,0.6,0}
\definecolor{gray}{rgb}{0.5,0.5,0.5}
\definecolor{mauve}{rgb}{0.58,0,0.82}

\lstdefinelanguage{QSharp}{%
    language     = Java,
    morekeywords = {namespace,open,operation,function,body,Qubit,let,mutable,repeat,until,fixup,newtype,fail,using,borrow,set,H,T,CNOT,adjoint,auto,controlled,Adjoint}
}

\title{Programming Quantum Computers \\ Using Design Automation}
\author{%
  \IEEEauthorblockN{Mathias Soeken$^1$ \qquad Thomas Haener$^2$ \qquad Martin Roetteler$^3$}
  \IEEEauthorblockA{%
    $^1$Integrated Systems Laboratory, EPFL, Lausanne, Switzerland \\
    $^2$Institute for Theoretical Physics, ETH Zurich, Switzerland\\
    $^3$Station Q, QuArC, Microsoft Research, Redmond, WA, USA
  }
}

\begin{document}
\maketitle

\begin{abstract}
  Recent developments in quantum hardware indicate that systems featuring more than 50 physical qubits are within reach. At this scale, classical simulation will no longer be feasible and there is a possibility that such quantum devices may outperform even classical supercomputers at certain tasks. With the rapid growth of qubit numbers and coherence times comes the increasingly difficult challenge of quantum program compilation. This entails the translation of a high-level description of a quantum algorithm to hardware-specific low-level operations which can be carried out by the quantum device.
  Some parts of the calculation may still be performed manually due to the lack of efficient methods. This, in turn, may lead to a design gap, which will prevent the
  programming of a quantum computer.
  In this paper, we discuss the challenges in fully-automatic quantum
  compilation.  We motivate directions for future research to tackle
  these challenges.  Yet, with the algorithms and approaches that
  exist today, we demonstrate how to automatically perform the quantum
  programming flow from algorithm to a physical quantum computer for a simple algorithmic benchmark, namely the hidden shift problem. We present and use two tool flows which invoke RevKit. One which is based on ProjectQ and which targets the IBM Quantum Experience or a local simulator, and one which is based on Microsoft's quantum programming language Q\#. 
\end{abstract}

\section{Introduction}
With the rapid development of quantum hardware, quantum computers will soon reach sizes---measured in the numbers of qubits on which they
operate---which allow them to solve problems that are out of reach for
any of the best classical supercomputers.  Quantum computers get this
computational advantage over classical computers from the principles of \emph{superposition} of states and \emph{interference} of computational paths. Arguably the most simple case in which superposition manifests itself is a single qubit which can be in any (normalized) linear combination of two basis states. In contrast, a bit in conventional computers is always in a single state. By linearly increasing the number of qubits, superposition allows quantum computers to exponentially increase their computational space, while still being able to execute operations on this exponentially large space at low cost (in concrete terms this means that, everything else being equal, e.g., a 17-qubit quantum computer is twice as powerful as a 16-qubit quantum computer). While classical probabilistic computation also allows to access an exponentially large space with only linear resources, a quantum computer can leverage the principle of interference which allows to amplify or reduce the probability of computational paths. When designed properly, a \emph{quantum algorithm} can combine the power to explore exponentially many computational paths at low cost with the ability to cancel out useless paths in such a way that a \emph{measurement} of the remaining paths reveals the answer to an interesting computational problem. 

Several quantum algorithms that are computationally superior to their
classical counterparts have already been found. The most prominent
one arguably is Shor's algorithm~\cite{Shor97} that allows to factorize
integers in polynomial time, whereas for classical computing nothing
better than a sub-exponential upper bound is known~\cite{Pomerance96}.
Consequently, Shor's algorithm can break public-key cryptography which
is based on the assumption that integer factorization is a hard task. Recently, precise cost estimates to implement Shor's algorithm for factoring \cite{HRS:2016} and elliptic curve dlog \cite{RNS+:2017} were obtained, based on implementing 
and testing large-scale Toffoli networks. 

In addition to Shor's algorithm, there are other quantum algorithms which play a role in scientific
applications of interest. Examples include: 
\begin{itemize}
\item Grover's search algorithm~\cite{Grover96}, which enables quadratically faster search in unstructured databases if the correct element can be recognized efficiently by a predicate (e.g., NP-complete problems). This has implications for the 
choice of security parameters in a post-quantum cryptographic world. 
Perhaps surprisingly, it turns out that the overhead due to implementing the defining predicate in a reversible way can be quite substantial \cite{GLR+:2016}. 
\item The HHL algorithm~\cite{HHL09}, which offers an exponential
  speedup for solving linear equations. Finding practical use cases of the HHL algorithm remains a challenge and the few real-world applications that have been identified so far~\cite{CJS13,SVM+:2017} require large overheads due to implementation of the classical subroutines that define the problem. 
\item Quantum simulation~\cite{Feynman:1982} (see, e.g.,~\cite{JCJ14,BCK:2015} for overviews and pointers to the literature) to model
  atomic-scale interactions efficiently \cite{ADL+:2005,WHW+:2015} allowing to approximate
  behavior in drugs, organics, and materials science \cite{SOG+:2002,BWM+:2016}, and has applications for simulating quantum field theories \cite{JLP:2012}.

\end{itemize}
In order to execute a quantum algorithm on a physical quantum computer, the algorithm must be expressed in terms of elementary \emph{quantum
  operations} that can be understood by a quantum computer---very much
like classical computer programs need to be expressed in terms of
low-level machine instructions to run on a classical computer.
\emph{Quantum compilers} are software programs that take a high-level
description of a quantum algorithm and map them into so-called
\emph{quantum circuits}.  Quantum circuits are \emph{not} a physical
entity, but an abstraction of the physical operations that can be
performed to qubits of the physical system~\cite{CFM+17}.  They are
represented in terms of sequences of low-level quantum operations.
Quantum circuits can be considered the ``assembly code'' of a quantum
computer, in which qubits play the role of registers.  The goal of
quantum compilers is to find a quantum circuit that meets the number
of available qubits and minimizes the number of quantum operations.
A challenge for quatum compilers is to map combinational non-quantum
operations into quantum circuits, while not exceeding the resource
constraints due to the limited number of qubits.  This is a 
difficult problem, and no satisfiable and sufficient solution is
provided by today's state-of-the-art quantum compilers.

Quantum computing has made a big leap this year, as research on
physical devices is moving from the academic environment into several
companies~\cite{Castelvecchi17}.  Microsoft, Google, IBM, Intel,
Alibaba, as well as the rapidly growing startup companies IonQ and
Rigetti, are investing into building the first
scalable quantum computer.  As of today, the largest publicly available
fully-programmable quantum computers\footnote{in contrast to
  special-purpose quantum computers such as the D-Wave quantum
  annealer} are by IBM which features 17
qubits~\cite{IBM17} and by Rigetti which features 19 qubits~\cite{Rigetti19}.  Recently, Intel announced a quantum
computer with 17 qubits~\cite{Intel17} and IBM quantum computers with up to 50 qubits~\cite{IBM17b}.  These sizes are
not yet practical, since it has been shown that supercomputers are
able to simulate low-depth quantum circuits with up to 56 qubits
classically~\cite{PGN+17}, and full state vector simulation is possible for up to 45 qubits~\cite{HS17}.  The rapid progress in quantum computing
and quantum simulation underlines the importance of having reliable
and robust quantum programming toolchains.

\section{Quantum Programming Languages}

Several quantum programming languages were proposed in recent years, ranging from imperative to functional and low-level to high-level~\cite{Miszczak2011}.  
Languages such as Quipper \cite{Selinger2013}, ScaffCC/Scaffold \cite{Scaffold,ScaffoldASPLOS}, LIQ$Ui|\rangle$ \cite{liquid}, QWire \cite{PRZ:2017}, Quil \cite{smith_practical_2016}, Q\# \cite{QSharp:2017,SGT+:2018} and ProjectQ \cite{SHT16} enable programming of quantum computers. Quipper is a strongly-typed, functional quantum programming language embedded in Haskell; Scaffold is a stand-alone C-like programming language and its compiler ScaffCC leverages the LLVM framework; QWire is embedded in the proof system Coq; LIQ$Ui|\rangle$ is embedded in F\#; Q\# is a stand-alone F\#-like language, and ProjectQ and Quil are embedded in Python.

All mentioned languages offer extensible frameworks for quantum circuit description and manipulation, and some of them offer gate decomposition and circuit optimization methods, some classical control flow, and exporting of quantum circuits for rendering or resource costing purposes. 

Theoretically, it would be sufficient if a programming language for quantum computing supported the gate set of the target hardware. The similarity between such an approach and classical assembly language brought into existence quantum assembly languages such as QASM~\cite{SAC+06} and OPENQASM~\cite{CBSG17}. While sufficient for today's quantum hardware which is able to perform a few gate operations on less than 20 qubits, programming in such a language is neither scalable nor particularly user-friendly. Rather, a quantum programming language should provide high-level abstractions in order to shorten development times and to enable portability across a wide range of quantum hardware backends, similar to today's compilers for classical high-level languages such as C$++$.

In addition to purely classical and purely quantum subroutines, typical quantum algorithms also require classical functions to be evaluated on a superposition of inputs, e.g., the modular exponentiation in Shor's algorithm for factoring~\cite{Shor97}. Therefore, such ``mixed'' constructs must also be supported by the language and the compiler must be able to translate these constructs to instructions which can be executed by the quantum hardware.

\section{Quantum computing basics}
This section introduces the necessary background on quantum algorithms
and quantum circuits.  This introduction is kept brief on purpose and
focuses on the most important notations and definitions that are
necessary in the course of this paper.  For a more detailed overview
on the matter, we refer the reader to the standard
literature~\cite{NC00}.

A \emph{quantum algorithm} is implemented in terms of a quantum
program, which is a sequence of high-level quantum operations that are
performed on a set of qubits.  A \emph{qubit state} is modeled as a
column
vector $|\varphi\rangle = \left(\begin{smallmatrix} \alpha_0 \\
    \alpha_1 \end{smallmatrix}\right)$ with two complex-valued
elements $\alpha_0$ and $\alpha_1$, called \emph{amplitudes}, such
that $|\alpha_0|^2 + |\alpha_1|^2 = 1$.  The values $|\alpha_0|^2$ and
$|\alpha_1|^2$ are the probabilities of whether the qubit state will
be 0 or 1 after measuring it, respectively.  The classical states for
a logic 0 and logic 1 are
$|0\rangle = \left(\begin{smallmatrix} 1 \\
    0 \end{smallmatrix}\right)$ and
$|1\rangle = \left(\begin{smallmatrix} 0 \\
    1\end{smallmatrix}\right)$, respectively.  Hence, we may also
write the state of a qubit as
$|\varphi\rangle = \alpha_0|0\rangle + \alpha_1|1\rangle$.  The
notation $|\cdot\rangle$ is called Dirac or \emph{bra-ket} notation
and typical for denoting quantum states.  A state in which the
measurement outcome has an equal probability of being 0 or 1 is for
example
$\frac{1}{\sqrt{2}}\left(\begin{smallmatrix} 1 \\
    1 \end{smallmatrix}\right)$, which is abbreviated as $|+\rangle$,
since it occurs very frequently in the design of quantum algorithms.
A different state with the same measurement probabilities is
$|-\rangle = \frac{1}{\sqrt{2}}\left(\begin{smallmatrix}1 \\
    -1\end{smallmatrix}\right)$.  Although the measurement probability
is the same, the quantum state is not, which is one reason that makes
quantum computing significantly different from probabilistic
computing.

Qubit registers refer to quantum states involving multiple qubits.  As
an example, a 2-qubit register is represented by the state
$\left(\begin{smallmatrix} \alpha_{00} \\ \alpha_{01} \\ \alpha_{10}
    \\ \alpha_{11} \end{smallmatrix}\right) = \alpha_{00}|00\rangle +
\alpha_{01}|01\rangle + \alpha_{10}|10\rangle +
\alpha_{11}|11\rangle$, which has four amplitudes, one for each
classical state $|00\rangle$, $|01\rangle$, $|10\rangle$, and
$|11\rangle$.  In general, an $n$-qubit register is a column vector
$|\varphi\rangle = \sum_{b \in \B^n}\alpha_b|b\rangle$ with $2^n$
amplitudes.  This reflects the exponential power of qubits.

\begin{figure}[t]
  \centering
  \subfloat[]{\raisebox{0.5cm}{\begin{tikzpicture}[scale=1.000000,x=1pt,y=1pt]
\filldraw[color=white] (0.000000, -7.500000) rectangle (42.000000, 22.500000);
\draw[color=black] (0.000000,15.000000) -- (42.000000,15.000000);
\draw[color=black] (0.000000,15.000000) node[left] {${|0\rangle}$};
\draw[color=black] (0.000000,0.000000) -- (42.000000,0.000000);
\draw[color=black] (0.000000,0.000000) node[left] {${|0\rangle}$};
\begin{scope}
\draw[fill=white] (12.000000, 15.000000) +(-45.000000:8.485281pt and 8.485281pt) -- +(45.000000:8.485281pt and 8.485281pt) -- +(135.000000:8.485281pt and 8.485281pt) -- +(225.000000:8.485281pt and 8.485281pt) -- cycle;
\clip (12.000000, 15.000000) +(-45.000000:8.485281pt and 8.485281pt) -- +(45.000000:8.485281pt and 8.485281pt) -- +(135.000000:8.485281pt and 8.485281pt) -- +(225.000000:8.485281pt and 8.485281pt) -- cycle;
\draw (12.000000, 15.000000) node {$H$};
\end{scope}
\draw (33.000000,15.000000) -- (33.000000,0.000000);
\filldraw (33.000000, 15.000000) circle(1.500000pt);
\begin{scope}
\draw[fill=white] (33.000000, 0.000000) circle(3.000000pt);
\clip (33.000000, 0.000000) circle(3.000000pt);
\draw (30.000000, 0.000000) -- (36.000000, 0.000000);
\draw (33.000000, -3.000000) -- (33.000000, 3.000000);
\end{scope}
\draw[color=black] (42.000000,7.500000) node[right] {${\ket{\Psi}}$};
\end{tikzpicture}}}
\hfil
  \subfloat[]{\begin{tikzpicture}[scale=1.000000,x=1pt,y=1pt]
\filldraw[color=white] (0.000000, -7.500000) rectangle (120.000000, 52.500000);
\draw[color=black] (0.000000,45.000000) -- (108.000000,45.000000);
\draw[color=black] (108.000000,44.500000) -- (120.000000,44.500000);
\draw[color=black] (108.000000,45.500000) -- (120.000000,45.500000);
\draw[color=black] (0.000000,45.000000) node[left] {${|\varphi_1\rangle}$};
\draw[color=black] (0.000000,30.000000) -- (108.000000,30.000000);
\draw[color=black] (108.000000,29.500000) -- (120.000000,29.500000);
\draw[color=black] (108.000000,30.500000) -- (120.000000,30.500000);
\draw[color=black] (0.000000,30.000000) node[left] {${|\varphi_2\rangle}$};
\draw[color=black] (0.000000,15.000000) -- (120.000000,15.000000);
\draw[color=black] (0.000000,15.000000) node[left] {${|\varphi_3\rangle}$};
\draw[color=black] (0.000000,0.000000) -- (120.000000,0.000000);
\draw[color=black] (0.000000,0.000000) node[left] {${|\varphi_4\rangle}$};
\begin{scope}
\draw[fill=white] (12.000000, 45.000000) +(-45.000000:8.485281pt and 8.485281pt) -- +(45.000000:8.485281pt and 8.485281pt) -- +(135.000000:8.485281pt and 8.485281pt) -- +(225.000000:8.485281pt and 8.485281pt) -- cycle;
\clip (12.000000, 45.000000) +(-45.000000:8.485281pt and 8.485281pt) -- +(45.000000:8.485281pt and 8.485281pt) -- +(135.000000:8.485281pt and 8.485281pt) -- +(225.000000:8.485281pt and 8.485281pt) -- cycle;
\draw (12.000000, 45.000000) node {{$R_1$}};
\end{scope}
\begin{scope}
\draw[fill=white] (12.000000, 15.000000) +(-45.000000:8.485281pt and 8.485281pt) -- +(45.000000:8.485281pt and 8.485281pt) -- +(135.000000:8.485281pt and 8.485281pt) -- +(225.000000:8.485281pt and 8.485281pt) -- cycle;
\clip (12.000000, 15.000000) +(-45.000000:8.485281pt and 8.485281pt) -- +(45.000000:8.485281pt and 8.485281pt) -- +(135.000000:8.485281pt and 8.485281pt) -- +(225.000000:8.485281pt and 8.485281pt) -- cycle;
\draw (12.000000, 15.000000) node {{$R_2$}};
\end{scope}
\draw (36.000000,30.000000) -- (36.000000,0.000000);
\begin{scope}
\draw[fill=white] (36.000000, 15.000000) +(-45.000000:8.485281pt and 29.698485pt) -- +(45.000000:8.485281pt and 29.698485pt) -- +(135.000000:8.485281pt and 29.698485pt) -- +(225.000000:8.485281pt and 29.698485pt) -- cycle;
\clip (36.000000, 15.000000) +(-45.000000:8.485281pt and 29.698485pt) -- +(45.000000:8.485281pt and 29.698485pt) -- +(135.000000:8.485281pt and 29.698485pt) -- +(225.000000:8.485281pt and 29.698485pt) -- cycle;
\draw (36.000000, 15.000000) node {{$U_1$}};
\end{scope}
\begin{scope}
\draw[fill=white] (60.000000, 30.000000) +(-45.000000:8.485281pt and 8.485281pt) -- +(45.000000:8.485281pt and 8.485281pt) -- +(135.000000:8.485281pt and 8.485281pt) -- +(225.000000:8.485281pt and 8.485281pt) -- cycle;
\clip (60.000000, 30.000000) +(-45.000000:8.485281pt and 8.485281pt) -- +(45.000000:8.485281pt and 8.485281pt) -- +(135.000000:8.485281pt and 8.485281pt) -- +(225.000000:8.485281pt and 8.485281pt) -- cycle;
\draw (60.000000, 30.000000) node {{$R_3$}};
\end{scope}
\begin{scope}
\draw[fill=white] (60.000000, -0.000000) +(-45.000000:8.485281pt and 8.485281pt) -- +(45.000000:8.485281pt and 8.485281pt) -- +(135.000000:8.485281pt and 8.485281pt) -- +(225.000000:8.485281pt and 8.485281pt) -- cycle;
\clip (60.000000, -0.000000) +(-45.000000:8.485281pt and 8.485281pt) -- +(45.000000:8.485281pt and 8.485281pt) -- +(135.000000:8.485281pt and 8.485281pt) -- +(225.000000:8.485281pt and 8.485281pt) -- cycle;
\draw (60.000000, -0.000000) node {{$R_4$}};
\end{scope}
\draw (84.000000,45.000000) -- (84.000000,15.000000);
\begin{scope}
\draw[fill=white] (84.000000, 30.000000) +(-45.000000:8.485281pt and 29.698485pt) -- +(45.000000:8.485281pt and 29.698485pt) -- +(135.000000:8.485281pt and 29.698485pt) -- +(225.000000:8.485281pt and 29.698485pt) -- cycle;
\clip (84.000000, 30.000000) +(-45.000000:8.485281pt and 29.698485pt) -- +(45.000000:8.485281pt and 29.698485pt) -- +(135.000000:8.485281pt and 29.698485pt) -- +(225.000000:8.485281pt and 29.698485pt) -- cycle;
\draw (84.000000, 30.000000) node {{$U_2$}};
\end{scope}
\draw[fill=white] (102.000000, 39.000000) rectangle (114.000000, 51.000000);
\draw[very thin] (108.000000, 45.600000) arc (90:150:6.000000pt);
\draw[very thin] (108.000000, 45.600000) arc (90:30:6.000000pt);
\draw[->,>=stealth] (108.000000, 39.600000) -- +(80:10.392305pt);
\draw[fill=white] (102.000000, 24.000000) rectangle (114.000000, 36.000000);
\draw[very thin] (108.000000, 30.600000) arc (90:150:6.000000pt);
\draw[very thin] (108.000000, 30.600000) arc (90:30:6.000000pt);
\draw[->,>=stealth] (108.000000, 24.600000) -- +(80:10.392305pt);
\end{tikzpicture}}
  \caption{Some basic examples for quantum circuits. Circuits are read from left to right. Shown in (a) is a simple quantum circuit that entangles two qubits. The circuit consists of a Hadamard gate $H$ and a controlled NOT gate and which creates upon input $\ket{0}\ket{0}$ the resulting output state $\ket{\Psi} = \frac{1}{\sqrt{2}}\left( \ket{00}+\ket{11}\right)$. Shown in (b) is  
an example for a larger quantum circuit consisting of local rotations $R_1, \ldots, R_4$ acting on single qubits, larger unitary gates $U_1, U_2$ acting on several qubits, and two measurement operations applied to the top two qubits.}

  \label{fig:qalg}
\end{figure}
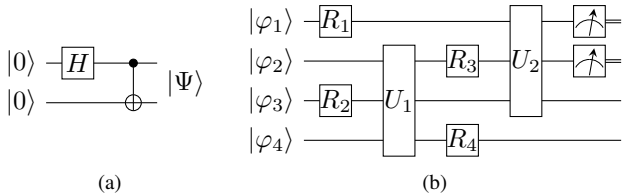

Quantum operations are modeled in terms of unitary matrices, called
\emph{quantum gates}.  A matrix $U$ is unitary if
$UU^{\dagger} = U^{\dagger}U = I$, where $U^{\dagger}$ refers to the
conjugate transpose of $U$ (also referred to as the Hermitian or
adjoint of $U$), and $I$ is the identity matrix.  A unitary matrix is
length-preserving and therefore maps one qubit state into another
qubit state.  A quantum operation that acts on a single qubit
is a $2 \times 2$ unitary matrix, and a quantum operation that
acts on an $n$-qubit register is a $2^n \times 2^n$ unitary
matrix.  An example for a single qubit operation is the so-called
\emph{Hadamard} operation
$H = \frac{1}{\sqrt{2}} \left(\begin{smallmatrix} 1 & 1 \\ 1 &
    -1 \end{smallmatrix}\right)$.  This operation can be used to
create a superposition of the two basis states $\ket0$ and $\ket1$, since $H|0\rangle = \frac 1{\sqrt 2}(\ket0 + \ket1)$.  The \emph{CNOT} operation is a 2-qubit
quantum operation that maps
$|\varphi_1\rangle|\varphi_2\rangle \mapsto
|\varphi_1\rangle|\varphi_1 \oplus \varphi_2\rangle$, where `$\oplus$'
is the exclusive-OR operation.  The CNOT operation inverts the
\emph{target qubit} $|\varphi_2\rangle$ if the \emph{control qubit}
$|\varphi_1\rangle$ is 1.  It can be represented as the unitary matrix
$\left(\begin{smallmatrix} 1 & 0 & 0 & 0 \\ 0 & 1 & 0 & 0 \\ 0 & 0 & 0
    & 1 \\ 0 & 0 & 1 & 0 \end{smallmatrix}\right)$.  The unitary
matrix of the CNOT operation is a permutation matrix.  A quantum
operation whose unitary matrix is a permutation matrix is called a
\emph{classical} operation.  This also means that all classical
operations must be reversible, since otherwise they cannot be
represented in terms of a permutation matrix.

A quantum algorithm describes the interaction of quantum operations
with qubits.  Researchers use \emph{quantum circuits} as an
abstraction to illustrate these interactions.  Fig.~\ref{fig:qalg}(a)
shows an abstract representation of such a quantum circuit.  The
horizontal lines represent qubits, the boxes represent quantum
operations that interact with the qubits, and time moves from left to right.  Consequently, the vertical
direction corresponds to space (i.e., number of qubits) and the
horizontal direction to time (i.e., number of quantum operations).
There are three types of operations: (i) quantum operations
($R_1, \dots, R_4$ in the figure), (ii) classical operations ($U_1$
and $U_2$ in the figure), and (iii) measurements which are illustrated
by a meter.  Classical operations perform classical computations, such
as arithmetic operations---but acting on qubits rather than bits.  A
quantum circuit can be seen as a way to represent a large unitary
matrix composed of smaller ones.  The absence of a gate in a circuit
corresponds to the identity matrix.  Fig.~\ref{fig:qalg}(b) shows a
simple quantum algorithm consisting of a Hadamard gate followed by a
CNOT operation.  The CNOT operation has a special notation with a
solid circle for the control qubit and an `$\oplus$' symbol for the
target qubit.  This quantum algorithm takes as input two qubits
initialized in state $|0\rangle$ and creates the 2-qubit state
$\frac{1}{\sqrt{2}}\left(|00\rangle + |11\rangle\right)$.  This state
is entangled, i.e., by measuring one qubit the outcome of the second
is immediately determined.  This also means that the explicit state of
one of the qubits cannot be described individually.  Sequential
composition of two gates in a quantum circuit corresponds to matrix
multiplication and parallel composition of gates corresponds to taking
the Kronecker product, denoted `$\otimes$'.  The unitary matrix
represented by the quantum circuit in Fig.~\ref{fig:qalg}(b) is
$\mathrm{CNOT}(H \otimes I)$.

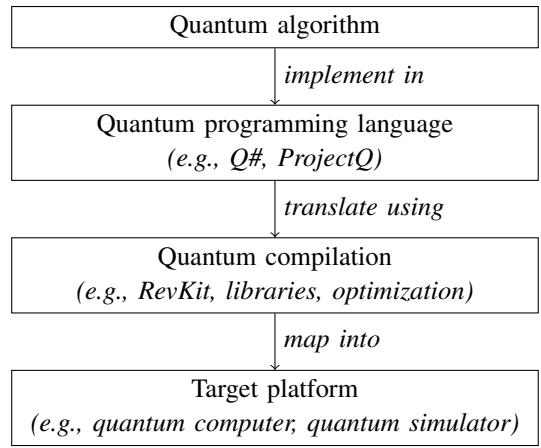
\begin{figure}[t]
  \centering
  \begin{tikzpicture}
  \begin{scope}[every node/.style={draw,minimum width=7cm}]
    \node (alg) {Quantum algorithm};
    \node[below=.75cm of alg,minimum height=1cm] (prog) {};
    \node[below=.75cm of prog,minimum height=1cm] (comp) {};
    \node[below=.75cm of comp,minimum height=1cm] (target) {};
  \end{scope}

  \node[anchor=north] at (prog.north) {Quantum programming language};
  \node[anchor=south] at (prog.south) {\textit{(e.g., Q\#, ProjectQ)}};

  \node[anchor=north] at (comp.north) {Quantum compilation};
  \node[anchor=south] at (comp.south) {\textit{(e.g., RevKit, libraries, optimization)}};

  \node[anchor=north] at (target.north) {Target platform};
  \node[anchor=south] at (target.south) {\textit{(e.g., quantum computer, quantum simulator)}};

  \draw[->] (alg.south) -- node[midway,right] {\textit{implement in}} (prog.north);
  \draw[->] (prog.south) -- node[midway,right] {\textit{translate using}} (comp.north);
  \draw[->] (comp.south) -- node[midway,right] {\textit{map into}} (target.north);
\end{tikzpicture}

  \caption{The high-level design flow for mapping quantum algorithms
    to quantum computers.}
  \label{fig:flow}
\end{figure}

Note that today's quantum algorithms rely on a variety of different
combinational calculations.  Factoring needs constant modular
arithmetic~\cite{Shor97}, computing elliptic curve discrete logarithms
using a quantum algorithm requires generic modular
arithmetic~\cite{RNS+:2017}, the HHL algorithm needs reciprocals and
Newton type methods~\cite{HHL09}, amplitude amplification algorithms
need implementations for search and collision~\cite{Grover96}, and
quantum simulation algorithms need addressing and indexing functions
for sparse matrices as well as computing Hamiltonian terms on the
fly~\cite{JCJ14}.

\section{Quantum design automation: general flow}
Fig.~\ref{fig:flow} abstractly illustrates the overall programming
flow for quantum computers.  The capabilities of the targeted quantum
computer are taken into account when developing the quantum algorithm.
A quantum algorithm consists of quantum parts and
classical combinational operations.  The quantum
algorithm must be translated into a quantum circuit.  While automatic
and satisfactory solutions exist for translating the quantum parts, no
sufficient solution exists for automatically translating the
combinational operations.  In fact, the current quantum programming
flow depends on predefined library components for which manually
derived quantum circuits exist.  Such a manual flow is time-consuming,
not flexible, and not scalable.

Rather, one would like to express the quantum program at a high level of abstraction and have a compiler which is able to automatically translate the entire circuit, even if no manually optimized libraries are available. It is thus crucial that the quantum programming language used to express the quantum program supports such a design flow.

\section{Compiling Boolean functions}
The translation of classical combinational operations into quantum
circuits involves \emph{reversible logic synthesis}~\cite{SM13}.  Due
to the physical properties of quantum states, all operations need to
be performed in a reversible manner.  State-of-the-art approaches
first create a reversible logic circuit with reversible gates, which
are Boolean abstractions of classical reversible operations.  Other
methods translate reversible gates into quantum
circuits~\cite{BBC+95,AASD16,Maslov16}.  Many approaches for
reversible logic synthesis have been proposed in the last 15
years~(e.g., \cite{MMD03,SPMH03,WD09,SWH+12}).

It is customary to distinguish reversible synthesis algorithms based
on whether the Boolean function that is input to the algorithm is
already a reversible function or not.  For a reversible Boolean
function $f : \B^n \to \B^n$, reversible synthesis algorithms find an
$n$ qubit quantum circuit that realizes the unitary
\begin{equation}
  \label{eq:reversible-mapping}
  U : |x\rangle \mapsto |f(x)\rangle.
\end{equation}
Several algorithms have been proposed for this task.  They differ
depending on $f$'s representation.  Most of the early algorithms
expect $f$ to be represented as a truth table (see,
e.g.,~\cite{MMD03,VR08,SZSS10,GWDD09b,MDM07}).  The explicit truth
table representation limits the application to large functions, i.e.,
$n > 20$.  Alternative implementations have been proposed that work on
symbolic representations of $f$, e.g., as binary decision diagrams
(BDDs)~\cite{SWH+12,STDD16} or Boolean satisfiability
problems~\cite{SDM16}.  These approaches are able to find quatum
circuits also for some Boolean functions that are much larger.
However, the symbolic function representation does not always
guarantee a compact function representation.  Nonetheless, the main
drawback of such reversible functions algorithms is that they require
a reversible input function, which is rarely the case in most
algorithms of interest.

The second class of reversible synthesis algorithms considers
irreversible functions $f : \B^n \to \B^m$ as input.  Since a quantum
circuit can not represent irreversible functions, $f$ must be embedded
into a reversible function.  This may be done either \emph{explicitly}
or \emph{implicitly}.  In the explicit case one finds a reversible
function $g : \B^r \to \B^r$ such that
\begin{equation}
  \label{eq:inplace-embedding}
  g(x_1, \dots, x_n, 0, \dots, 0) = (y_1, \dots, y_m, y_{m+1}, \dots,
y_r),
\end{equation}
if $f(x_1, \dots, x_n) = (y_1, \dots, y_m)$.  Finding $g$ such that
$r$ is minimum is coNP-hard~\cite{SWK+16} and does therefore not scale
to larger functions, although symbolic methods can help to slightly
increase the range of applicable functions~\cite{SWK+16,ZW17}.  An
embedding as in~\eqref{eq:inplace-embedding} is referred to as
\emph{in-place embedding}, since the input values are not restored
after the application of $g$.  As an example, explicit embedding with
symbolic reversible logic synthesis was applied to find in-place
reversible circuits for the reciprocal function $1/x$ up to
$n = m = 16$ digits in $x$ and $r = 31$ (see~\cite{SRWM17}).

One can easily show that there exists a reversible function $g$ with
$r = m + n$, by chosing
\begin{equation}
  \label{eq:bennett}
  g(x, y) = (x, y \oplus f(x))
\end{equation}
where $x = x_1, \dots, x_n$, $y = y_1, \dots, y_m$, and `$\oplus$'
refers to the bitwise application of the XOR operation in this case.
Such an embedding is also referred to as \emph{Bennett embedding}.
So-called ESOP (exclusive sum-of-products) based reversible synthesis
approaches~\cite{FTR07,MP02,BRD14} find reversible circuits that
realize~\eqref{eq:bennett}.  In order to apply ESOP-based synthesis
one must find 2-level ESOP expressions for each of the $m$ outputs in
$f$ (see, e.g., \cite{Drechsler99,MP01}).  This approach can be time
consuming and limits the application to large functions.
In~\cite{SRWM17} ESOP-based synthesis was successfully applied up to
$n = 25$ for the reciprocal function.

Scalable reversible synthesis algorithms require additional helper
qubits, called ancillae.  Given an irreversible Boolean function
$f: \B^n \to \B^m$, they find an $(n+m+k)$ qubit quantum circuit that
realizes the unitary
\begin{equation}
  \label{eq:ancilla-synthesis}
  U : |x\rangle|y\rangle|0^k\rangle \mapsto |x\rangle|y \oplus f(x)\rangle|0^k\rangle.
\end{equation}
If $k=0$, the synthesis problem corresponds to ESOP-based synthesis,
but for $k > 0$, the synthesis algorithm can use the $k$ additional
qubits to store intermediate computations.  The most effective methods
use multi-level logic network representations such as
BDDs~\cite{WD09,SWD10,CLA+15}, And-inverter
graphs~\cite{SC16,Valiron16}, XOR-majority graphs~\cite{SRWM17}, or
LUT networks~\cite{SRWM17b}.  These methods are referred to as
\emph{hierarchical} reversible logic synthesis.  Intermediate results
represented by internal nodes in the corresponding logic networks are
mapped on the additional qubits.  If the network has many internal
nodes, many ancillae are required, however, pebbling
strategies~\cite{Kralovic01} may be employed to trade off the number
of qubits for quantum operations~\cite{PRS17}.  One of the biggest
problems in hierarchical reversible logic synthesis is the fact that
$k$ is a result of the synthesis algorithm, i.e., it is determined by
the algorithm's requirements for temporary storage.  One of the
largest challenges in reversible logic synthesis is to find reversible
synthesis algorithms that take $k$ as an input parameter and guarantee
to return a quantum circuit that satisfies the space requirements.

\section{Illustrative Example}
In this and the following sections, we use the example of the hidden shift problem for
Boolean functions to illustrate the complete flow of programming a
quantum computer.  For this purpose, we leverage the quantum
programming languages ProjectQ~\cite{SHT16} and Q\#~\cite{QSharp:2017} interfaced with the
quantum compilation framework RevKit~\cite{SFWD12}.  ProjectQ and Q\# allow
for a high-level description of the algorithm using several
meta-constructs, and enables interfacing a physical quantum computer.
RevKit is used to automatically translate the combinational parts in
the quantum algorithm for the hidden shift problem into quantum gates.

ProjectQ is an open source software framework for quantum computing with a modular compiler design which allows domain experts to easily extend its functionality. Furthermore, this modularity enables portability of quantum algorithm implementations. Specifically, once an algorithm has been implemented, it can be run using various types of backends, be it software (simulator, emulator, resource counter, etc.) or hardware (classical and/or quantum).

Q\# \cite{QSharp:2017} is a scalable, multi-paradigm, domain-specific programming language for quantum computing by Microsoft. The Q\# framework allows describe how instructions are executed on quantum machines. The machines that can be targeted include many different levels of abstraction, ranging from various simulators to actual quantum hardware. Q\# is multi-paradigm in that it supports functional and imperative programming styles. Q\# is scalable in that it allows to write programs to target machines of various sizes, ranging from small machines with only a few hundred qubits to large machines with millions of qubits. Finally, being a bona-fide stand-alone language, Q\# allows a programmer to code complex quantum algorithms, offers rich and informative error reporting, and allows to perform various tasks such as debugging, profiling, resource estimation, and certain special-purpose simulations.

RevKit is an open source C++ framework and library that implements a
large set of reversible synthesis, optimization, and mapping
algorithms.  By default, RevKit is executed as a command-based shell
application, which allows to perform synthesis scripts by combining a
variety of different commands.  As an example, the command sequence
\begin{equation}
  \label{eq:revkit-cmds}
  \footnotesize
  \text{\texttt{revgen -{}-hwb 4; tbs; revsimp; rptm; tpar; ps -c}}
\end{equation}
generates a reversible function describing the 4-input reversible
hidden-weighted bit function, synthesizes it into a reversible circuit
using transformation-based synthesis~\cite{MMD03}, performs
simplification of the resulting circuit, maps it into Clifford+$T$
gates using the mapping described in~\cite{Maslov16}, optimizes the
$T$ count using the T-par algorithm presented in~\cite{AMM14}, and
finally prints statistics about the final quantum circuit.  All RevKit
commands provided by the shell can also be accessed via a Python
interface, e.g., `\texttt{revkit.revgen(hwb = 4)}' for the first
command in~\eqref{eq:revkit-cmds}.  Using the Python interface, RevKit
can be executed from within ProjectQ using the
\texttt{projectq.libs.revkit} module.

\subsection{Quantum algorithm: the Boolean hidden shift problem}
For the illustrative example, we review the hidden shift problem for
Boolean functions~\cite{Rotteler10}. In general, the hidden shift
problem is a quite natural source of problems for which a quantum
computer might have an advantage over a classical computer as it
exploits the property that fast convolutions can be performed by
computing Fourier transforms and pointwise
multiplication. See~\cite{CD10} for general background on hidden
shifts and related problems and~\cite{Rotteler10} for the case of
hidden shifts over Boolean functions. Recently, the hidden shift
problem for bent functions was also studied in~\cite{BG16} from the
point of view of classical simulation of the resulting quantum
circuits.  The problem of computing hidden shifts for Boolean
functions is the following:

\begin{definition}[Hidden shift problem]
  Let $n \geq 1$ and let $f,g : \B^n \rightarrow \B$ be two Boolean
  functions such that the following conditions hold: (i) $f$, and $g$
  are bent functions, and (ii) there exist $s\in \B^n$ such that
  $g(x)=f(x+s)$ for all $x\in \B^n$. Moreover, let oracle access for
  $g$ and the dual bent function $\widetilde{f}$ be given. The task is
  then to find the hidden shift $s$.
\end{definition}

Bent functions are Boolean functions which have a perfectly flat
Fourier (i.e., Hadamard) spectrum, which in a sense makes them resemble random noise. It is easy to see that bent functions can only exist if the number of variables $n$ is even. What makes the hidden shift problem for bent
functions attractive is that it can be shown that classical algorithms
cannot find the shift efficiently, whereas quantum algorithms can find
the shift with only $1$ query to $g$ and $1$ query to
$\widetilde{f}$. Moreover, the quantum algorithm to find hidden shifts
is very simple as shown in Fig.~\ref{fig:algs}: the gates needed are
Hadamard gates, diagonal unitaries to implement the shifted function
and the dual bent function, and measurements in the computational
basis. An attractive feature of the algorithm is that---assuming
perfect gates---the answer is deterministic, i.e., the measured bit
pattern directly corresponds to the hidden shift.  We assign each
operation in the quantum algorithm an index from 1 to 6, written below
each gate.

\subsection{Maiorana-McFarland bent functions}
Arguably, the most simple example for a bent function is the inner
product $f(x,y) = xy^t = \sum_{i=1}^nx_iy_i$ of two bit-vectors
$x = x_1, \dots, x_n$ and $y = y_1, \dots, y_n$. Note that this is a Boolean function $f:\B^{2n}\to\B$ on an even number $2n$ of variables. The function can be
generalized to
\begin{equation}
  \label{eq:maiorana-mcfarland}
  f(x, y) = x\pi(y)^t + h(y)
\end{equation}
for an arbitrary permutation $\pi \in S_{2^n}$ on all $2^n$ boolean bitvectors of length $n$ and an
arbitrary Boolean function $h : \B^n \to \B$.  This leads to the class
of so-called Maiorana-McFarland bent functions\footnote{Named after mathematicians James A.~Maiorana (1946--2014) and Robert L.~McFarland who were the first to study these functions about 50 years ago.}.  The dual bent
function is
$\widetilde{f}(x,y) = \pi^{-1}(x)y^t +
h(\pi^{-1}(x))$~\cite{Rotteler10}. Asymptotically, the size of this class scales as $O(2^{c n 2^n})$ which is doubly exponential in $n$, however, which is just an exponentially small fraction of the set of all Boolean functions on $2n$ variables. A simple counting argument shows that most permutation $\pi$ do not have an efficient circuit, however, there exist natural families of Maiorana-McFarland bent function for which the permutation $\pi$ as well as the Boolean function $h$ can be implemented efficiently. 

\begin{figure}[t]
\centering
\begin{tikzpicture}[scale=1.000000,x=1pt,y=1pt]
\filldraw[color=white] (0.000000, -7.500000) rectangle (183.000000, 7.500000);
\draw[color=black] (0.000000,0.000000) -- (159.000000,0.000000);
\draw[color=black] (159.000000,-0.500000) -- (183.000000,-0.500000);
\draw[color=black] (159.000000,0.500000) -- (183.000000,0.500000);
\draw[color=black] (0.000000,0.000000) node[left] {${|0\rangle^{\otimes n}}$};
\draw (2.000000, -6.000000) -- (10.000000, 6.000000);
\draw (26.500000, -7.500000) node[text width=144pt,below,text centered] {{\footnotesize 1}};
\begin{scope}
\draw[fill=white] (26.500000, 0.000000) +(-45.000000:17.677670pt and 12.020815pt) -- +(45.000000:17.677670pt and 12.020815pt) -- +(135.000000:17.677670pt and 12.020815pt) -- +(225.000000:17.677670pt and 12.020815pt) -- cycle;
\clip (26.500000, 0.000000) +(-45.000000:17.677670pt and 12.020815pt) -- +(45.000000:17.677670pt and 12.020815pt) -- +(135.000000:17.677670pt and 12.020815pt) -- +(225.000000:17.677670pt and 12.020815pt) -- cycle;
\draw (26.500000, 0.000000) node {{$H^{\otimes n}$}};
\end{scope}
\draw (53.000000, -7.500000) node[text width=144pt,below,text centered] {{\footnotesize 2}};
\begin{scope}
\draw[fill=white] (53.000000, 0.000000) +(-45.000000:14.142136pt and 12.020815pt) -- +(45.000000:14.142136pt and 12.020815pt) -- +(135.000000:14.142136pt and 12.020815pt) -- +(225.000000:14.142136pt and 12.020815pt) -- cycle;
\clip (53.000000, 0.000000) +(-45.000000:14.142136pt and 12.020815pt) -- +(45.000000:14.142136pt and 12.020815pt) -- +(135.000000:14.142136pt and 12.020815pt) -- +(225.000000:14.142136pt and 12.020815pt) -- cycle;
\draw (53.000000, 0.000000) node {{$U_g$}};
\end{scope}
\draw (79.500000, -7.500000) node[text width=144pt,below,text centered] {{\footnotesize 3}};
\begin{scope}
\draw[fill=white] (79.500000, 0.000000) +(-45.000000:17.677670pt and 12.020815pt) -- +(45.000000:17.677670pt and 12.020815pt) -- +(135.000000:17.677670pt and 12.020815pt) -- +(225.000000:17.677670pt and 12.020815pt) -- cycle;
\clip (79.500000, 0.000000) +(-45.000000:17.677670pt and 12.020815pt) -- +(45.000000:17.677670pt and 12.020815pt) -- +(135.000000:17.677670pt and 12.020815pt) -- +(225.000000:17.677670pt and 12.020815pt) -- cycle;
\draw (79.500000, 0.000000) node {{$H^{\otimes n}$}};
\end{scope}
\draw (106.000000, -7.500000) node[text width=144pt,below,text centered] {{\footnotesize 4}};
\begin{scope}
\draw[fill=white] (106.000000, 0.000000) +(-45.000000:14.142136pt and 12.020815pt) -- +(45.000000:14.142136pt and 12.020815pt) -- +(135.000000:14.142136pt and 12.020815pt) -- +(225.000000:14.142136pt and 12.020815pt) -- cycle;
\clip (106.000000, 0.000000) +(-45.000000:14.142136pt and 12.020815pt) -- +(45.000000:14.142136pt and 12.020815pt) -- +(135.000000:14.142136pt and 12.020815pt) -- +(225.000000:14.142136pt and 12.020815pt) -- cycle;
\draw (106.000000, 0.000000) node {{$U_{\tilde f}$}};
\end{scope}
\draw (132.500000, -7.500000) node[text width=144pt,below,text centered] {{\footnotesize 5}};
\begin{scope}
\draw[fill=white] (132.500000, 0.000000) +(-45.000000:17.677670pt and 12.020815pt) -- +(45.000000:17.677670pt and 12.020815pt) -- +(135.000000:17.677670pt and 12.020815pt) -- +(225.000000:17.677670pt and 12.020815pt) -- cycle;
\clip (132.500000, 0.000000) +(-45.000000:17.677670pt and 12.020815pt) -- +(45.000000:17.677670pt and 12.020815pt) -- +(135.000000:17.677670pt and 12.020815pt) -- +(225.000000:17.677670pt and 12.020815pt) -- cycle;
\draw (132.500000, 0.000000) node {{$H^{\otimes n}$}};
\end{scope}
\draw (159.000000, -7.500000) node[text width=144pt,below,text centered] {{\footnotesize 6}};
\draw[fill=white] (149.000000, -8.500000) rectangle (169.000000, 8.500000);
\draw[very thin] (159.000000, 0.850000) arc (90:150:8.500000pt);
\draw[very thin] (159.000000, 0.850000) arc (90:30:8.500000pt);
\draw[->,>=stealth] (159.000000, -7.650000) -- +(80:14.722432pt);
\draw (173.000000, -6.000000) -- (181.000000, 6.000000);
\draw[color=black] (183.000000,0.000000) node[right] {${|s\rangle}$};
\end{tikzpicture}
\caption{\label{fig:algs} Quantum algorithm for the hidden shift
  problem for a bent functions $f$. The quantum circuit assumes access
  to the shifted function $g(x) = f(x+s)$ which is implemented by the
  diagonal unitary operator $U_g = \sum_x (-1)^{g(x)}
  \ket{x}\bra{x}$. Also, the algorithm needs access to the dual bent
  function $\widetilde{f}$, which again is computed into the phase via
  a diagonal unitary.}
\end{figure}
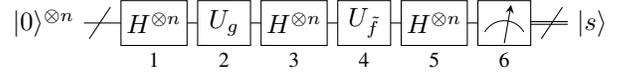

The same basic circuit as shown in Fig.~\ref{fig:algs} can be used to solve the hidden shift problem for Maiorana-McFarland bent functions. Note however, that in contrast to the case of the inner product function for which $\widetilde{f}=f$ holds, for the more general case where $\pi$ is not the identity permutation, the diagonal unitary $U_f = \sum_{x} (-1)^{f(x)} \ket{x}\bra{x}$ implementing the bent function $f$ and the shift $g$ is different from the diagonal unitary $U_{\widetilde{f}} = \sum_{x} (-1)^{{\widetilde{f}}(x)} \ket{x}\bra{x}$ implementing the dual bent function. For the Maiorana-McFarland family specifically, the difference in implementing $f$ and $\widetilde{f}$ is to use the inverse permutation $\pi^{-1}$ instead of $\pi$ and to apply it to the $x$-variables instead of the $y$ variables and similarly the role of $x$ and $y$ has to be changed in the evaluation of $h$.  

\section{Interop with ProjectQ and simulator / IBM backend}

In this section, we show how to program a concrete instance of the
hidden shift problem using ProjectQ and RevKit.  We choose
$f(x) = x_1 x_2 \oplus x_3 x_4$ as a Boolean function on $4$
variables, and $g(x) = f(x + 1)$, i.e., $s = 1$.  It can be shown that
$f=\widetilde{f}$.

\begin{figure}[t]
\begin{lstlisting}[language=Python,basicstyle=\footnotesize\ttfamily,breaklines=true,numbers=left,numberstyle=\tiny\color{gray},numbersep=5pt,showspaces=false,xleftmargin=10pt,keywords={from,import,def,with,return,and},escapechar=@]
from projectq.cengines import MainEngine
from projectq.ops import All, H, X, Measure
from projectq.meta import Compute, Uncompute
from projectq.libs.revkit import PhaseOracle

# phase function
def f(a, b, c, d):
    return (a and b) ^ (c and d)

eng = MainEngine() @\label{l1:init_b}@
x1, x2, x3, x4 = qubits = eng.allocate_qureg(4) @\label{l1:init_e}@

# circuit
with Compute(eng): @\label{l1:compute}@
    All(H) | qubits @\label{l1:step1}@
    X | x1 @\label{l1:shift}@
PhaseOracle(f) | qubits @\label{l1:phase}@
Uncompute(eng) @\label{l1:unc}@

PhaseOracle(f) | qubits @\label{l1:phaseinv}@
All(H) | qubits @\label{l1:step5}@
Measure | qubits @\label{l1:meas}@

eng.flush()

# measurement result
print("Shift is {}".format(8 * int(x4) + 4 * int(x3) + 2 * int(x2) + int(x1)))
\end{lstlisting}
\caption{ProjectQ python code for an instance of the hidden shift
  problem where $f(x) = x_1x_2 \oplus x_3x_4$ and $g(x) = f(x + 1)$.}
\label{fig:code1}
\end{figure}

\begin{figure}[t]
  \centering
  \input{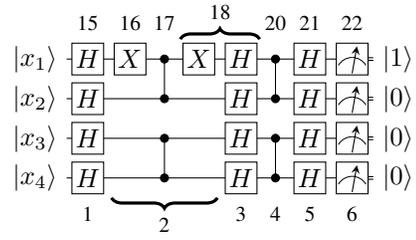}
  \caption{Quantum circuit that is implemented by the Python code in Fig.~\ref{fig:code1}; indexes below the gates correspond to the steps in Fig.~\ref{fig:algs}, indexes above the gates correspond to the lines in Fig.~\ref{fig:code1}.}
  \label{fig:code1-qc}
\end{figure}

\begin{figure}[t]
  \centering
  \resizebox{\linewidth}{!}{\input{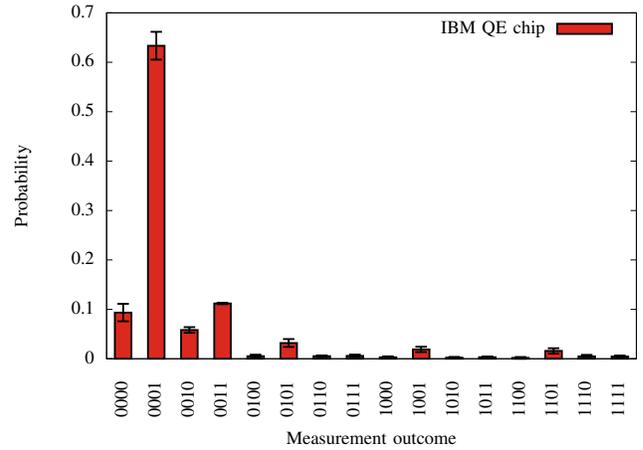}}
  \caption{Histogram depicting the average and standard deviation of the outcome probabilities of three runs of the code in Fig.~\ref{fig:code1}. Each run consists of 1024 executions of the circuit on the IBM Quantum Experience chip. The correct shift $s=1$ was found with average probability $p\approx 0.63$.}
  \label{fig:ibmhist}
\end{figure}

Fig.~\ref{fig:code1} shows the ProjectQ Python code for this example.
The corresponding quantum circuit that is generated by the code is
shown in Fig.~\ref{fig:code1-qc}.  Lines
\ref{l1:init_b}--\ref{l1:init_e} initialize a ProjectQ engine with 4
qubits, named \texttt{x1}, \texttt{x2}, \texttt{x3}, and \texttt{x4},
and stored in a list \texttt{qubits}.  Line~\ref{l1:step1} performs
step~1 of the quantum algorithm described in Fig.~\ref{fig:algs}.
Line~\ref{l1:shift} describes the shift by $s = 1$, implemented using
an $X$ operation on the least-significant qubit \texttt{x1}.  Together
with the phase circuit computed for $f$ in line~\ref{l1:phase}, it
computes step 2 in the quantum algorithm.  As input to the
\texttt{PhaseOracle} statement we can provide a predicate \texttt{f}
implemented as Python function.  The \texttt{PhaseOracle} statement
converts the Python code in \texttt{f} into a Boolean expression.
This expression is then passed to RevKit, which automatically compiles
the expression into a circuit computing the function described by
\texttt{f} into the global phase of the circuit.  The
\texttt{Uncompute} statement in line~\ref{l1:unc} uncomputes all
operations that were specified in the \texttt{Compute} block in
lines~\ref{l1:compute}--\ref{l1:shift}, by applying all operations in
inverse order.  This will also add step 3 of the algorithm to the
quantum circuit.  Since $\widetilde{f} = f$, we again compute the
phase circuit for \texttt{f} in line~\ref{l1:phaseinv}, apply Hadamard
gates to each qubit for step 5 of the algorithm, and finally measure
all qubits in line~\ref{l1:meas}.  The resulting state of the qubits,
computed using simulation, corresponds to the shift $s=1$.  The
program outputs `\texttt{Shift is 1}.' By changing two lines of code in~\ref{fig:code1}, the backend can be changed to the IBM Quantum Experience chip. Doing so and running three times 1024 shots of the circuit yielded the results depicted in Fig.~\ref{fig:ibmhist}.

\begin{figure}[t]
\begin{lstlisting}[language=Python,basicstyle=\footnotesize\ttfamily,breaklines=true,numbers=left,numberstyle=\tiny\color{gray},numbersep=5pt,showspaces=false,xleftmargin=10pt,keywords={from,import,def,with,return,and},escapechar=@]
from projectq.cengines import MainEngine
from projectq.ops import All, H, X, Measure
from projectq.meta import Compute, Uncompute, Dagger
from projectq.libs.revkit import PhaseOracle, PermutationOracle
import revkit

# phase function
def f(a, b, c, d, e, f):
  return (a and b) ^ (c and d) ^ (e and f)

# permutation
pi = [0, 2, 3, 5, 7, 1, 4, 6]

eng = MainEngine()
qubits = eng.allocate_qureg(6)
x = qubits[::2]  # qubits on odd lines
y = qubits[1::2] # qubits on even lines

# circuit
with Compute(eng):
  All(H) | qubits
  All(X) | [x[0], x[1]]
  PermutationOracle(pi) | y
PhaseOracle(f) | qubits
Uncompute(eng)

with Compute(eng):
  with Dagger(eng):
    PermutationOracle(pi, synth=revkit.dbs) | x
PhaseOracle(f) | qubits
Uncompute(eng)

All(H) | qubits
Measure | qubits

eng.flush()

# measurement result
print("Shift is {}".format(sum(int(q) << i for i, q in enumerate(qubits))))
\end{lstlisting}
\caption{ProjectQ python code for an instance of the hidden shift
  problem where $f(x, y) = x\pi(y)^t$, $\pi = [0, 2, 3, 5, 7, 1, 4, 6]$, and $s = 5$.}
\label{fig:code2}
\end{figure}

\begin{figure*}[t]
  \centering
  \input{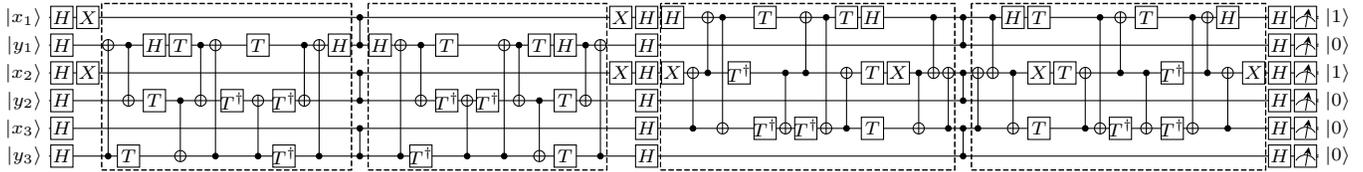}
  \caption{Quantum circuit that is implemented by the Python code in
    Fig.~\ref{fig:code2}.  The dashed boxes emphasize the subcircuits
    which correspond to realizations of $\pi$ and its inverse.}
  \label{fig:code2-qc}
\end{figure*}

Fig.~\ref{fig:code2} shows a Python code that implements an instance
of the hidden shift problem for a Maiorana-McFarland bent function
where $n=3$, $\pi = [0, 2, 3, 5, 7, 1, 4, 6]$, and $h = 0$.
Fig.~\ref{fig:code2-qc} shows the corresponding circuit.  The program
is similar to the program in Fig.~\ref{fig:code1}.  We create 6 qubits
and partition them into three qubits \texttt{x} for $x_1, x_2, x_3$
and three qubits \texttt{y} for $y_1, y_2, y_3$.  The inner product is
realized by the bent function specified by the Python function
\texttt{f}.  We use the function \texttt{PermutationOracle} to create
a quantum circuit from a permutation, which is then applied to the
qubits in \texttt{x}.  The function \texttt{PermutationOracle} calls
RevKit using transformation-based synthesis~\cite{MMD03} followed by a
mapping of Toffoli gates into Clifford+$T$ gates using the algorithm
presented in~\cite{Maslov16}.  For the second part of the circuit, we
need a quantum circuit for the inverse permutation $\pi^{-1}$.
Instead of inverting $\pi$, we compute another quantum circuit for
$\pi$ and invert the circuit using the \texttt{Dagger} statement.
Note that for this compilation we chose decomposition-based
synthesis~\cite{VR08} for finding a Toffoli network for the
permutation.  Since each permutation is uncomputed after the phase
circuit for the inner product, the final circuit consists of four
subcircuits realizing either $\pi$ or its inverse.  These are
emphasized using dashed boxes in Fig.~\ref{fig:code2-qc}.

\section{Interop with Q\# and simulator backend}

In the following we describe a programming flow that implements the same 
high-level algorithm, i.e., an instance of the hidden shift problem 
for Maiorana-McFarland functions, but implements it in Q\#. While at a high level, the interop between RevKit and Q\# happens as
described in Fig.~\ref{fig:flow}, the actual invocation of RevKit in the design flow is slightly different from the RevKit/ProjectQ interop in that RevKit is used as a pre-processor to produce the code for the permutation oracle as Q\# native code. Subsequently, the Q\# compiler is then invoked to compile the algorithm and to target a simulator backend that is part of the Microsoft Quantum Development Kit (QDK). 

\begin{figure}[t]
\lstset{numbers=left}
\begin{lstlisting}[basicstyle=\footnotesize\ttfamily]
namespace Microsoft.Quantum.HiddenShift{
// basic operations: Hadamard, CNOT, etc
open Microsoft.Quantum.Primitive;
// useful lib functions and combinators
open Microsoft.Quantum.Canon;
// permutation defining the instance
open Microsoft.Quantum.PermOracle;

operation HiddenShift 
// signature of input types 
(Ufstar : (Qubit[] => ()), 
Ug : (Qubit[] => ()), n : Int) : 
// signature of output type
Result[] {
    body {
        mutable resultArray = new Result[n];
        // allocate n clean qubits
        using(qubits=Qubit[n]) {				
            ApplyToEach(H, qubits);
            Ug(qubits);
            ApplyToEach(H, qubits);
            Ufstar(qubits);
            ApplyToEach(H, qubits);
            // measure and reset qubits
            for (idx in 0..(n-1)) {
                set resultArray[idx] = MResetZ(qubits[idx]);
            }
        }	
        Message($"result: {resultArray}");
        return resultArray; 
     }
}}
\end{lstlisting}
\caption{Implementation of the correlation algorithm for the Boolean hidden shift problem in Q\#. This code is shipped as an algorithm sample with the Microsoft QDK \cite{QSharp:2017}.}
\label{fig:code3}
\end{figure}

The code for the hidden shift problem shown in Fig.~\ref{fig:code3} is structurally quite similar to familiar languages such as C\# and Java in its use of semicolons to end statements, curly brackets to group statements, and double-slash to introduce comments. Q\# also uses namespaces to group definitions together, and allows references to elements from other namespaces.

The Q\# code begins with a namespace statement (line 1) which declares the symbols and makes then available for other projects. The mechanism to include other namespaces is via the \texttt{open} keyword. This is used here in line 3 to include the basic gates such as the Hadamard gate $H$ and in line 5 to include the ``canon'' which is a large library of useful operations, functions, and combinators. For the current example we use operations \texttt{ApplyToEach} and \texttt{MResetZ} from the canon. The implementation of the permutation oracle itself is provided in another namespace which is included in line 7. The basic unit in Q\# to model side effects on quantum data is an operation such as the operation \texttt{HiddenShift} declared in line 9. Besides operations, Q\# also supports functions which allow to modify state that is purely classical. Note that the definition of an operation or a function must begin with a declaration of the type signature of the function, including its input and output types. This is done in lines 10--14 of the present example. 

Operations and functions are first-class citizens in Q\#, i.e., they can be passed as arguments. In the present case, \texttt{Ustar} is an operation that implements the diagonal operator $U_{\widetilde{f}}$ as defined earlier. If an operation changes the state of a quantum register (modeled here as \texttt{Qubit[]} array), then its type is \texttt{Qubit[] => ()}, where \texttt{()} denotes the unit type. 
Operations are the only way the state of an abstract quantum machine model can be manipulated. Q\# can be used to target many abstract quantum machine models, including future physical implementations of scalable quantum computers. Currently, the main target of Q\# is a state-of-the-art simulator that can easily handle up to 30 qubits on a standard computer and over 40 qubits on a distributed computer using an MPI-based implementation. 

The \texttt{body} element on line 15 specifies the implementation of the operation.
Q\# operations may also specify implementations for variants, or derived operations, that are common in quantum computing. These variants indicated by \texttt{adjoint} (inverse), a \texttt{controlled} and \texttt{controlled adjoint}. If the key-word \texttt{auto} is provided, then the compiler automatically calculates the inverse or controlled version of the operation based on the body, but in general it can make sense to provide these implementations separately as more efficient circuits might be known. While variants do not occur in the implementation of the \texttt{HiddenShift} operation, they do occur in the implementation of the operation \texttt{PermutationOracle} further below. 

Q\# allows the introduction of mutable variable as in line 16 which is returned to a driver program (which can be written in a .NET language such as C\# or F\#) in line 30. Further notable elements used in this code snippet are the allocation of clean qubits (which by definition are initialized in the $\ket{0}$ state) in line 18 by using the \texttt{using} keyword. Q\# offers classical flow and control constructs like in line 25 where the code iterates through a range of integers using \texttt{for}. Finally, we remark that Q\# supports mutable and immutable types. The syntax for declaring a new \texttt{mutable} variable is shown in line 16 of an array that will hold the 
final result of the computation. Assignment of mutable variables is done using  \texttt{set} statements as in line 26. 

\begin{figure}[t]
\lstset{numbers=left}
\begin{lstlisting}[basicstyle=\footnotesize\ttfamily]
namespace Microsoft.Quantum.PermOracle{
open Microsoft.Quantum.Primitive;

operation PermutationOracle
// signature of input types 
(qubits : Qubit[]) : 
// signature of output type
() {
    body {
        CNOT(qubits[2], qubits[1]);
        H(qubits[0]);
        T(qubits[2]);
        T(qubits[1]);
        T(qubits[0]);
        CNOT(qubits[1], qubits[2]);
        CNOT(qubits[0], qubits[1]);
        CNOT(qubits[2], qubits[0]);
        (Adjoint T)(qubits[1]);
        CNOT(qubits[2], qubits[1]);
        (Adjoint T)(qubits[2]);
        (Adjoint T)(qubits[1]);
        T(qubits[0]);
        CNOT(qubits[0], qubits[1]);
        CNOT(qubits[2], qubits[0]);
        CNOT(qubits[1], qubits[2]);
        H(qubits[0]);
        CNOT(qubits[0], qubits[1]);
        CNOT(qubits[1], qubits[2]);
    }
    adjoint auto
    controlled auto
    controlled adjoint auto
}

operation BentFunctionImpl
(n : Int, qs : Qubit[]) : () {
    body {
       let xs = qs[0..(n-1)]; 
		let ys = qs[u..(2*n-1)];
		(Adjoint PermutationOracle)(ys); 
		for (idx in 0..(n-1)) {
			(Controlled Z)([xs[idx]], ys[idx]);
		}
		PermutationOracle(ys);
	}
}

function BentFunction
(n : Int) : (Qubit[] => ()) {
	return BentFunctionImpl(n, _); 		
}}
\end{lstlisting}
\caption{Q\# code for an instance of the hidden shift
  problem where $f(x, y) = x\pi(y)^t$, $\pi = [0, 2, 3, 5, 7, 1, 4, 6]$.}
\label{fig:code4}
\end{figure}

The definition of the instance $U_g$ and $U_{\widetilde{f}}$ of the hidden shift problem itself is done by calling RevKit first during a pre-processing state. The input for this is a description of the permutation $\pi$ to be implemented. The output of this stage is another Q\# program which is shown as the \texttt{PermutationOracle} operation in Fig.~\ref{fig:code4}. Note that this operation makes use of primitive gates that are built-into the Q\# language and that are native to the underlying abstract quantum machine model, such as $H$, $T$, and ${\rm CNOT}$. Also note that the \texttt{Adjoint} functor is used in lines 18, 20, and 21 which computes the inverse of the invoked operation. 

The instance of the bent function is defined in the block starting at line 48 and returns a function with signature \texttt{Qubit[] => ()}. The implementation of this function, which depends on the number of variables (here denoted by integer \texttt{n} using the \texttt{Int} primitive type) invokes another operation from which the function is constructed using partial application, which is the basic mechanism in which e.g. currying can be implemented in Q\#. 

For space reasons, not all subroutines used in the implementation of the shifted bent functions and the test harness are shown as snippets, however, these can be inspected as sample Q\# code that was shipped with the QDK \cite{QSharp:2017}. The test subroutine consists of a C\# part that invokes the above Q\# program and targets the built-in simulator.  

\section{Challenges and Conclusions}
In this paper, we illustrated and discussed the high-level design flow
for mapping a quantum algorithm to quantum computers using quantum
programming languages.  Expressive syntactical constructs and a rich
API in combination with effective automatic compilation algorithms
allow us to express quantum algorithms at a high level without being
burdened with specifying each single quantum operation.  This
ultimately leads to implement (i) more scalable algorithms, since
tedious manual compilation of combinational components is performed
automatically, and (ii) more complex algorithms by combining abstract
high-level syntactic constructs offered by the programming language.
Therefore, programming quantum computers is catching up with
its classical counterpart in which a variety of high-level
programming languages and significant effort in the development of
compilers render manual assembly descriptions unnecessary.

Several challenges remain and are awaiting satisfactory solutions.  In
this paper, we only considered simple reversible
synthesis methods which do not require additional ancilla qubits for
the realization of the quantum circuit.  This limits their application
to small functions with up to about 25 variables.  In order to
automatically compile larger functions, reversible logic synthesis
methods require additional qubits.  These are typically determined
during the execution of the algorithm, and cannot be bounded ahead of time.
Synthesis methods that find a solution without exceeding a given number of ancillae are rare and the state of available solutions is
still in its infancy~\cite{SRWM17b,PRS17}.

Another issue is the verification of the synthesized circuits.  Simulating
the quantum circuit may require to represent the complete quantum
state, which is exponentially large in the number of qubits.  Verified compilers that are
``correct-by-construction'' address this issue~\cite{ARS16}.  However,
when applying post-optimization, one needs to verify that the optimized
circuit did not change the functionality, requiring to simulate
complete quantum states in the worst-case.

\section*{Acknowledgments}

We thank the QuArC team for useful discussions. Circuits were typeset using $\langle q | pic \rangle$ by Tom Draper and Sandy Kutin.



\end{document}